\documentstyle[11pt,moriond,epsfig,psfrag]{article}
\def\met{\mbox{${\hbox{$E$\kern-0.6em\lower-.1ex\hbox{/}}}_T$}} 
\def\metx{\mbox{${\hbox{$E$\kern-0.6em\lower-.1ex\hbox{/}}}_x$}} 
\def\mety{\mbox{${\hbox{$E$\kern-0.6em\lower-.1ex\hbox{/}}}_y$}} 
\def\metcal{\mbox{${\hbox{$E$\kern-0.6em\lower-.1ex\hbox{/}}}_T^{cal}$}} 
\def\mez{\mbox{${\hbox{$E$\kern-0.6em\lower-.1ex\hbox{/}}}_z$}} 
\def\gev{GeV}                           
\def\D0{D\O}                            
\def\etal{{\sl et al.}}                 

\newcommand{\PRL}{{\em Phys. Rev. Lett.} }

\def\err#1#2#3 {{\it Erratum} {\bf#1},{\ #2} (19#3)}
\def\ib#1#2#3 {{\it ibid.} {\bf#1},{\ #2} (19#3)}
\def\nc#1#2#3 {Nuovo Cim. {\bf#1} ,#2(19#3)}
\def\nim#1#2#3 {Nucl. Instr. Meth. {\bf#1},{\ #2} (19#3)}
\def\np#1#2#3 {Nucl. Phys. {\bf#1},{\ #2} (19#3)}
\def\pl#1#2#3 {Phys. Lett. {\bf#1},{\ #2} (19#3)}
\def\prev#1#2#3 {Phys. Rev. {\bf#1},{\ #2} (19#3)}
\def\prd#1#2#3 {Phys. Rev. D {\bf#1},{\ #2} (19#3)}
\def\prb#1#2#3 {Phys. Rev. B {\bf#1},{\ #2} (19#3)}
\def\prl#1#2#3 {Phys. Rev. Lett. {\bf#1},{\ #2} (19#3)}
\def\rmp#1#2#3 {Rev. Mod. Phys. {\bf#1},{\ #2} (19#3)}
\def\zp#1#2#3 {Zeit. Phys. {\bf#1},{\ #2} (19#3)}

\def\Journal#1#2#3#4{{#1} {\bf #2}, #3 (#4)}

\def\NIMA{{\em Nucl. Instrum. Methods} A}
\def\NPB{{\em Nucl. Phys.} B}
\def\PLB{{\em Phys. Lett.}  B}
\def\PRL{\em Phys. Rev. Lett.}
\def\PRD{{\em Phys. Rev.} D}
\def\ZPC{{\em Z. Phys.} C}

\def\be{\begin{equation}}
\def\ee{\end{equation}}
\def\bea{\begin{eqnarray}}
\def\eea{\end{eqnarray}}

\begin{document}
\vspace*{4cm} \title{NEW QCD RESULTS FROM NUTEV}

\author{M.~Tzanov$^7$, T. Adams$^4$, A.~Alton$^4$,
  S.~Avvakumov$^8$, L.~de~Barbaro$^5$, P.~de~Barbaro$^8$,
  R.~H.~Bernstein$^3$, A.~Bodek$^8$, T.~Bolton$^4$, S.~Boyd$^7$, 
  J.~Brau$^6$, D.~Buchholz$^5$, H.~Budd$^8$, L.~Bugel$^3$, J.~Conrad$^1$,
  R.~B.~Drucker$^6$, B.~T. Fleming$^1$, J.~Formaggio$^1$, R.~Frey$^6$,
  J.~Goldman$^4$, M.~Goncharov$^4$, D.~A.~Harris$^8$, R.~A.~Johnson$^2$, 
  J.~H.~Kim$^1$, S.~Koutsoliotas$^1$, M.~J.~Lamm$^3$,
  W.~Marsh$^3$, D.~Mason$^6$, J.~McDonald$^7$, 
  K.~S.~McFarland$^8$, C.~McNulty$^1$,
  D.~Naples$^7$, P.~Nienaber$^3$, V.~Radescu$^7$, A.~Romosan$^1$,
  W.~K.~Sakumoto$^8$,H.~Schellman$^5$, M.~H.~Shaevitz$^1$,
  P.~Spentzouris$^1$, E.~G.~Stern$^1$, N.~Suwonjandee$^2$,
  N.~Tobien$^3$, A.~Vaitaitis$^1$, M.~Vakili$^2$,
  U.~K.~Yang$^8$, J.~Yu$^3$, G.~P.~Zeller$^5$,
  E.~D.~Zimmerman$^1$\\
  {\it The NuTeV Collaboration}}

\address{ \it \baselineskip=13pt $^1$Columbia University, New York,
  NY, $^2$University of Cincinnati, Cincinnati, OH, $^3$Fermi National
  Accelerator Laboratory, Batavia, IL, $^4$Kansas State University,
  Manhattan, KS, $^5$Northwestern University, Evanston, IL,
  $^6$University of Oregon, Eugene, OR, $^7$University of Pittsburgh,
  Pittsburgh, PA, $^8$University of Rochester, Rochester, NY.}

\maketitle \abstracts{ Preliminary results from next-to-leading order
   strange sea fits to the NuTeV's deep inelastic scattering
   charged-current dimuon cross section data are presented. 
   Preliminary neutrino and antineutrino inclusive charged-current 
   differential cross sections and
   preliminary structure function $F_2(x,Q^2)$ are presented.}

\section*{Introduction} 

Neutrino-nucleon deep inelastic scattering (DIS) probes 
the structure of the nucleon and QCD. The NuTeV experiment is a 
high-energy fixed target $\nu N$ 
scattering experiment~\cite{nuRev}, which combines two new features:
separate high-purity neutrino and antineutrino beams, used to 
tag the primary lepton in charged-current interactions, 
and a continuous precision calibration beam which improves
the experiment's knowledge of the absolute energy scale
for hadrons and muons produced in neutrino interactions~\cite{nutCal}.
 $8.6\times10^5$ $\nu$ and $2.3\times10^5$ $\overline{\nu}$ 
charged-current (CC) interactions passing analysis cuts
were collected during NuTeV's data taking run.

\section*{NLO Charm Production}

Charm production occurs in CC interactions when
 $\nu$($\overline{\nu})$ scatters off $s(\overline{s})$ or
$d(\overline{d})$ quarks in the nucleon. 
Charm production from $d$ quarks is Cabbibo suppressed,
which allows sensitivity to the strange parton distribution function
(PDF). The charmed hadron produced in this 
interaction decays semileptonically to a muon about $10 \%$
of the time providing a  distinct signature of two 
final state muons of opposite sign.
The high-purity
beam tags the sign of the muon from the primary vertex and 
allows an independent measurement of strange and antistrange seas.

The presence of a large gluon sea in the nucleon requires
the strange sea fit to be performed at NLO in QCD. A new NLO calculation
\cite{nlo1}
implemented into a FORTRAN routine DISCO~\cite{disco}, 
calculates the differential 
cross section for charm production as differential in $x$, $y$, 
fragmentation momentum ratio $z$, 
and charm rapidity $\eta_c$.

NuTeV recently published a measurement of the differential 
cross section for forward dimuon production~\cite{dicross}, 
with the muon from charm decay, required to 
have energy $E_{\mu 2}>5$ \gev.
A NLO fit to the forward dimuon cross section is performed using,
\begin{equation}
  \frac{d^2\sigma_{2\mu}}{dxdy} = \frac{d^2\sigma_{charm}}{dxdy}\times
  EMC \times B_c \times A
\label{eq:dfe1}
\end{equation}
 where $\frac{d^2\sigma_{2\mu}}{dxdy}$ is the measured NuTeV dimuon
cross section, $\frac{d^2\sigma_{charm}}{dxdy}$ is the NLO cross 
section model integrated over $z$ and $\eta_c$, $EMC$ is an $x$ 
dependent correction for the EMC effect
and nuclear shadowing, $B_c$ is the semileptonic charm branching
ratio ($B_c=0.093$~\cite{Bc} is used in the fit), and $A$ is an acceptance
correction accounting for the minimum energy cut on $E_{\mu 2}$.
The acceptance correction is a function of the fragmentation 
parameter $\epsilon$ and the charm mass $m_c$,
in addition to $x$, $y$, and $E_\nu$.
Both cross sections are functions of neutrino energy $E_\nu$.

\subsection*{Preliminary Strange Sea Fit Results}

\begin{figure}
\psfig{file= 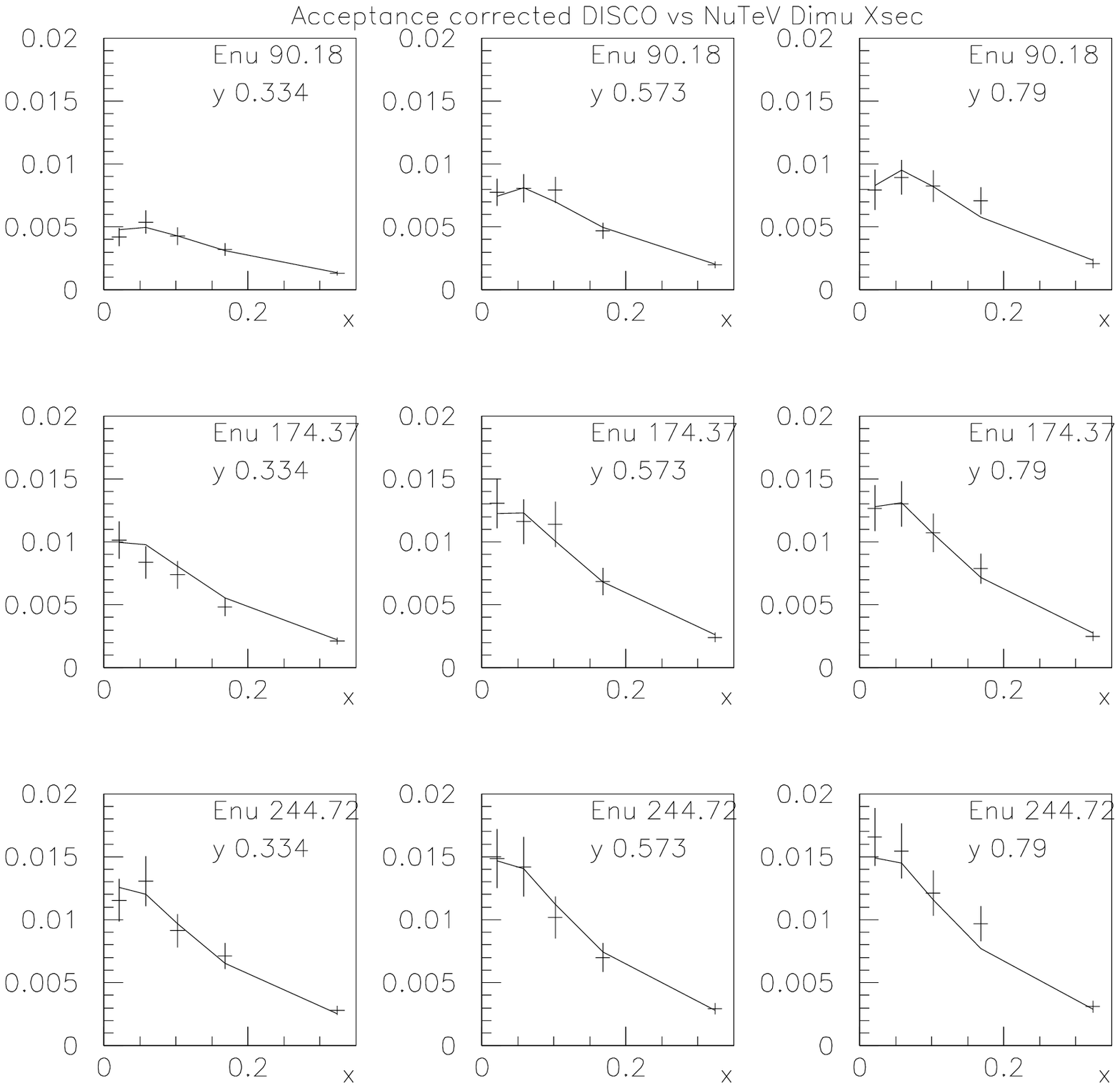, height=7cm}
\psfig{file= 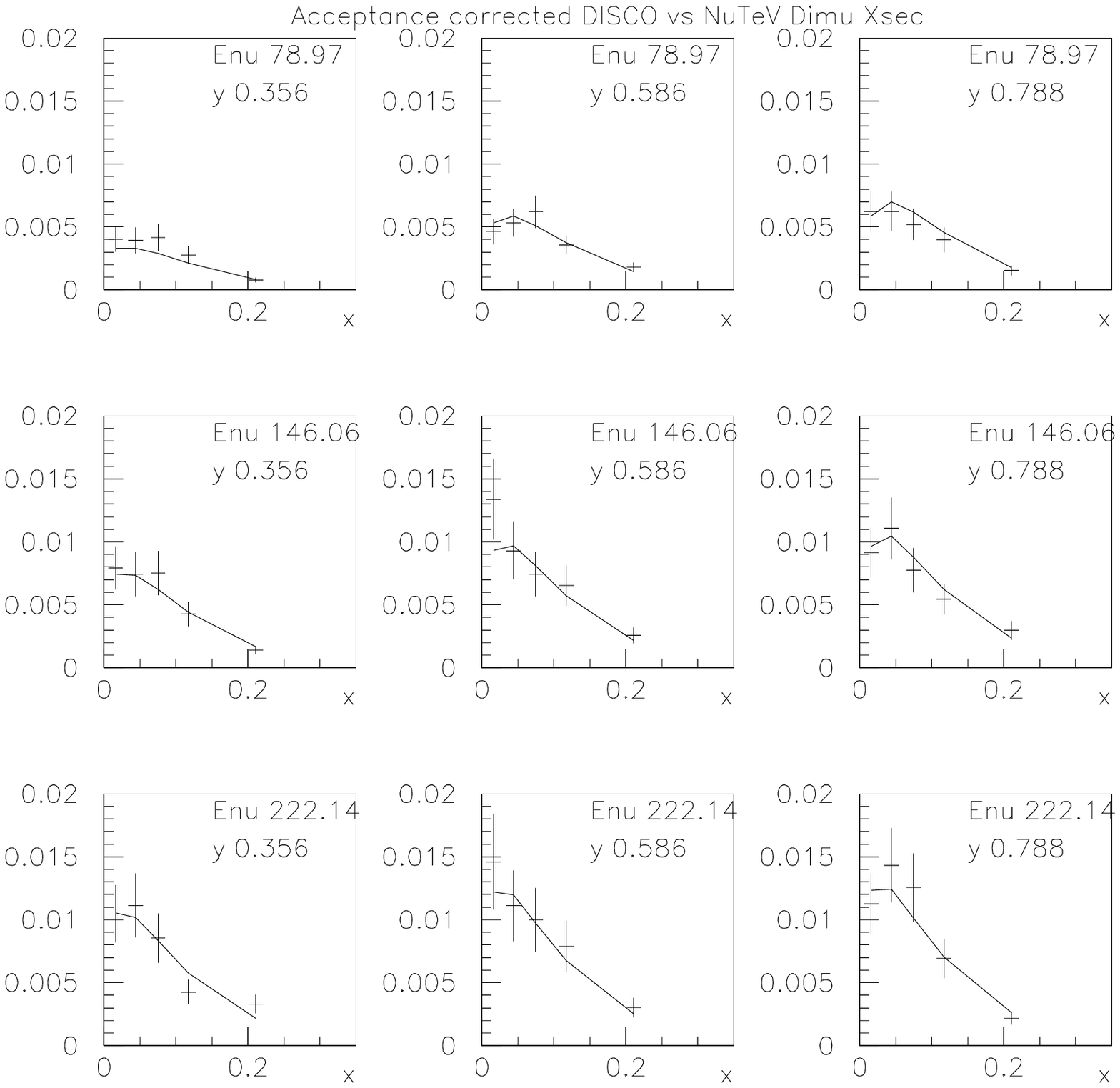, height=7cm}
\caption{NuTeV's forward dimuon cross section (crosses) with the 
NLO fit curve for $\nu$ (left) and $\overline{\nu}$ (right).}
\label{fig:charmfit}
\end{figure}

A preliminary fit to the NuTeV dimuon cross section table has been 
performed using the NLO cross section model described in the previous
section. 
\begin{table}[t]
\begin{center}
\begin{tabular}{ |l|l|l|l| } \hline
parameter & fit value & stat. & syst. \\ \hline
$m_c$ & $1.46$ & $\pm 0.24$ & $\pm 0.07$ \\ \hline
$\kappa$ & $0.516$ & $\pm 0.033$ & $\pm 0.031$ \\ \hline
$\overline \kappa$ & $0.511$ & $\pm 0.038$ & $\pm 0.040$ \\ \hline
$\alpha$ & $0.73$ & $\pm 0.47$ & $\pm 0.52$ \\ \hline
$\overline \alpha$ & $0.92$ & $\pm 0.43$ & $\pm 0.06$ \\ \hline
$\epsilon$ & $0.20$ & $\pm 0.13$ & $\pm 0.04$ \\ \hline
$\chi^2/dof$ &     $36.2/38$ &   &   \\ \hline
\end{tabular}
\caption {Preliminary 6-parameter fit to the NuTeV's
dimuon cross section table.}
\end{center}
\label{tab:sfit}
\end{table}
Collins-Spiller fragmentation parameter $\epsilon$, charm mass $m_c$, 
$\kappa$, $\alpha$, $\overline{\kappa}$, and $\overline{\alpha}$ are
the 6 fit parameters. The $s(\overline s)$ sea is parameterized 
using,
\begin{equation}
  s(\overline s) = \kappa (\overline \kappa)
  {\frac{\overline{u}(x)+\overline{d}(x)}{2}}(1-x)^{\alpha (\overline \alpha)}
\label{eq:sea}
\end{equation}
in terms of the light-quark sea distributions. 
The GRV94HO \cite{grv} PDF's were used in the fit.

Fig.~\ref{fig:charmfit} shows the preliminary fit result and the
dimuon cross section points for both neutrino and antineutrino 
mode. Fit parameters are given in Table~1.
The NLO fit describes the data well. This result is the first 
NLO extraction with full differential dependence 
on $\eta_c$ and $z$. 
This NLO model will be used to 
re-extract the dimuon cross section as a check.
 
\section*{$\nu$-Fe CC Differential Cross Section and
$\nu$-Fe Structure Functions}
 
The differential cross section is determined from
\begin{equation}
  \frac{d^2\sigma^{\nu(\overline{\nu})}}{dxdy} =  \frac{1}{\Phi(E)}\frac{d^2 N^{\nu(\overline{\nu})}(E)}{dxdy}. 
\label{eq:ccxsec}
\end{equation}
where $\Phi(E)$ is the $\nu (\overline{\nu})$ flux in energy bins. 
The cross section event sample is required to pass fiducial volume 
cuts, $\mu$ track reconstruction quality cuts, and minimum 
energy thresholds for muon energy ($E_\mu>15$~GeV), hadronic energy 
($E_{\scriptstyle HAD}>10$~GeV), and neutrino energy ($E_\nu>30$~GeV).
Selected events are binned in $x$, $y$, and $E_\nu$
bins, and corrected for acceptance and smearing using a 
detector simulation. 
$Q^2>1$~GeV$^2$/c$^2$ is required 
to minimize the non-perturbative contribution to the cross section.
NuTeV data ranges from $10^{-3}$ to $0.95$ in $x$,
$0.05$ to $0.95$ in y, and from 30~GeV to 300~GeV in $E_\nu$.

The flux is determined from data with $E_{\scriptstyle
HAD}<20$ \gev\ using the ``fixed $\nu_0$'' relative flux 
extraction method~\cite{nuRev}. 
The integrated number of events
in this sample as $y=\frac{E_{\scriptscriptstyle HAD}}{E_\nu}\rightarrow 0$
is proportional to the flux. Corrections, determined from 
the data sample, up to order $y^2$
are applied to determine the relative flux to about the $1\%$ level. 
Flux is normalized using the world average $\nu$-Fe cross
section~\cite{pdg} $\frac{\sigma^\nu}{E_\nu}=0.677 \times 10^{-38} cm^2/GeV$.

\begin{figure} 
\hspace{1.3cm}\begin{minipage}[t]{.45\textwidth}
\parbox[t]{0.45\textwidth}{
\psfrag{unitsTag}[][][1][90]{$\frac{1}{E_{\nu}}
\frac{d^2\sigma}{dxdy} ~\times {\scriptstyle 10}^{-38}$~{\tiny cm$^2$/GeV)}}
\psfrag{yaxisTag}[][][1]{$\scriptstyle Y$}
\includegraphics[width=7.0cm,height=6.3cm]{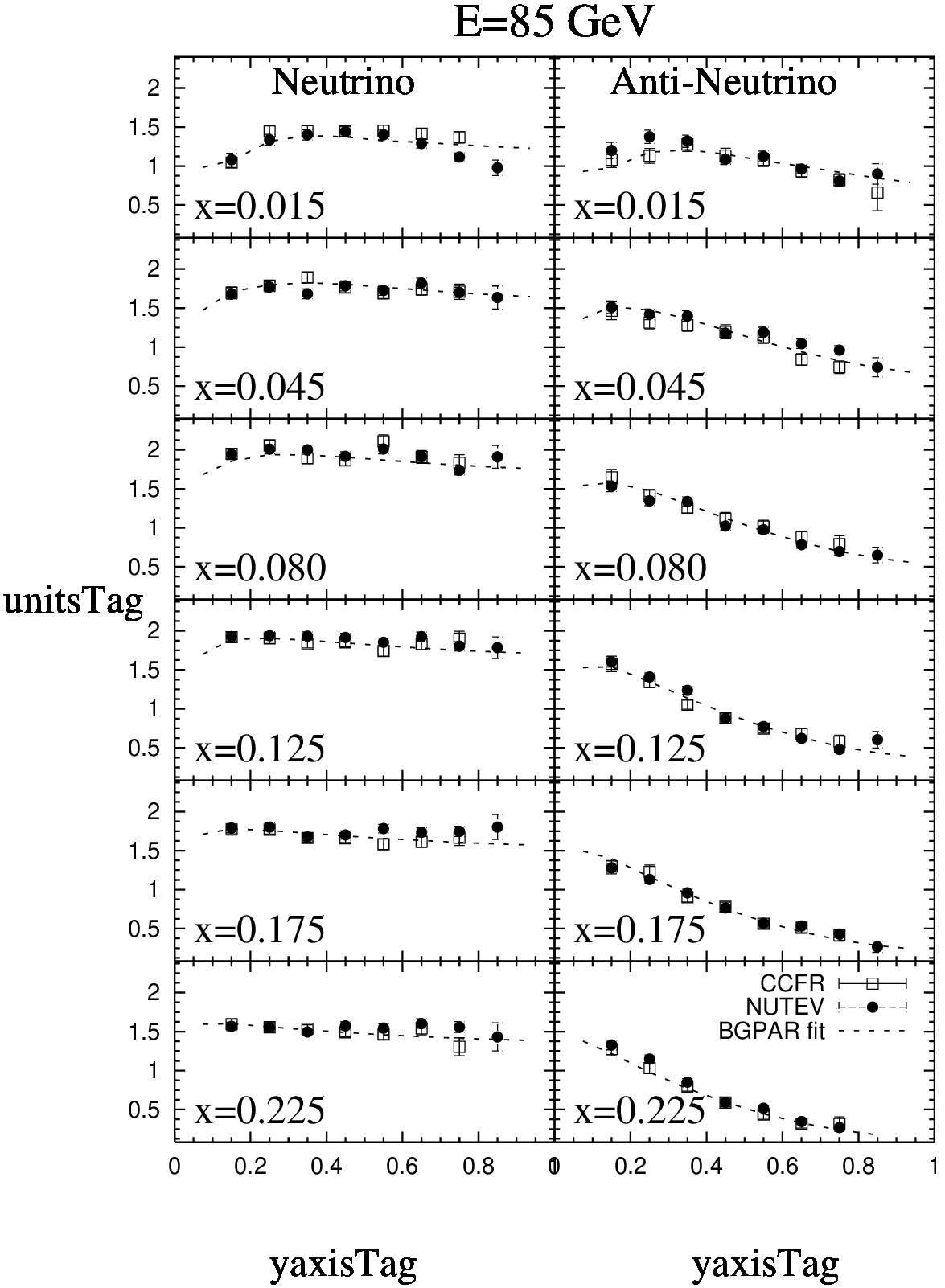}}
\end{minipage} 
\begin{minipage}[t]{.45\textwidth}
\psfig{figure=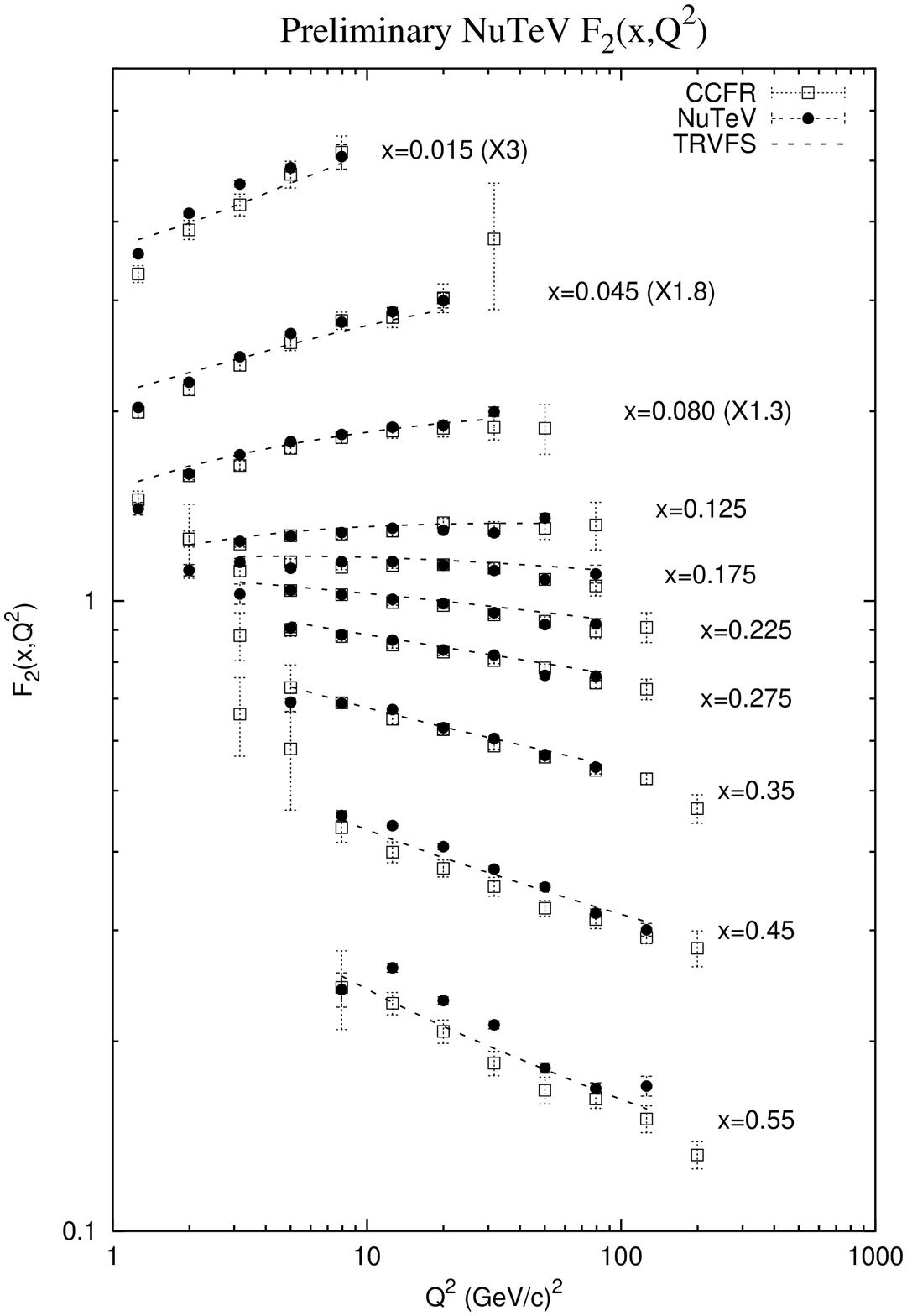,width=5.7cm,height=6.3cm}
\end{minipage} 
\caption{(Left) Preliminary cross section measurement
NuTeV (solid squares), CCFR (open squares), curve shows 
QCD inspired fit to data. 
(Right) Preliminary $F_2$ measurement NuTeV (solid squares), CCFR
(open squares), curve shows NLO QCD Model$^{12}$. 
Statistical errors only.}
\label{fig:dfxs}
\end{figure}

The detector simulation, which takes into account acceptance and
resolution effects, uses an empirically determined set of PDFs
extracted by fitting the differential cross section~\cite{bg}. 
The procedure is then iterated until convergence is achieved (within 3
iterations). 
Detector response functions are parameterized from the
NuTeV calibration beam data samples~\cite{nutCal}.
 As a typical example, the
differential cross section at $E = 85$ GeV is shown on
Fig.~\ref{fig:dfxs}(left).  NuTeV is found to be in good agreement with
CCFR~\cite{ccfr}.

\subsection*{Preliminary $F_2$ Structure Function Fit Result}

The sum of the neutrino and antineutrino differential cross sections 
can be written in terms of $\epsilon$, the polarization of W boson :
\begin{equation}
   F(\epsilon) =  \frac{\pi (1-\epsilon)}{y^2 G_F^2 M_p E_{\nu}}
      \left( \frac{d^2\sigma^\nu}{dxdy} +
        \frac{d^2\sigma^{\overline{\nu}}}{dxdy} \right) 
 =  2xF_1\left[1+\epsilon R(x,Q^2)\right] + 
  \frac{y(1-\frac{y}{2})}{1+(1-y)^2+M_{p}xy/E} \Delta xF_3.
\label{eq:feps2} 
\end{equation}
where $\epsilon = \frac{2(1-y)- M_pxy/E}{1+(1-y)^2 +  M_pxy/E}$.
$F_2(x,Q^2)$ is extracted from fits to the $y$ distribution 
using $R_{WORLD}$~\cite{Rworld} as input for $R(x,Q^2)$,
and a NLO model~\cite{trvfs} for $\Delta xF_3$.
Preliminary $F_2$ fit result is shown on Fig.~\ref{fig:dfxs} (right).  
Results are in good agreement with previous $\nu-Fe$ measurements
for $x<0.45$ and good agreement with QCD.

\section*{Conclusions and Prospects}

NuTeV has new preliminary results on 
NLO charm production from fits to the NuTeV dimuon cross section. 
NLO charm mass parameter and $s$($\overline{s}$) seas have been 
extracted from the fits. 
New preliminary measurements of the $\nu$-Fe and $\overline{\nu}$-Fe
CC differential cross sections
and the structure function $F_2$ have been presented. 
Both are in good agreement with the results from previous $\nu$-Fe 
experiments and QCD.

\end{document}